\documentclass[superscriptaddress,twocolumn,showpacs,amsmath,amssymb]{revtex4}

\usepackage{graphicx}

\begin{document}

\title{
Unveiling pairing anti-halo effect in odd-even staggering 
in reaction cross sections of weakly bound nuclei}

\author{K. Hagino}
\affiliation{ 
Department of Physics, Tohoku University, Sendai, 980-8578,  Japan} 

\author{H. Sagawa}
\affiliation{
Center for Mathematics and Physics,  University of Aizu, 
Aizu-Wakamatsu, Fukushima 965-8560,  Japan}



\begin{abstract}
We investigate the spatial extension of weakly bound Ne and C isotopes 
by taking into account the pairing correlation with 
the Hartree-Fock-Bogoliubov (HFB) method 
and a 3-body model, respectively. 
We show that the odd-even staggering in the reaction cross sections 
of $^{30,31,32}$Ne and $^{14,15,16}$C 
are successfully reproduced, and thus the staggering can be attributed 
to the pairing anti-halo effect. 
A correlation  between a one-neutron separation energy 
and the anti-halo effect
is demonstrated for 
$s$- and $p$-waves  
using the HFB wave functions.
\end{abstract}


\maketitle

Interaction cross sections  $\sigma_I$ 
and reaction cross sections $\sigma_R$ 
of unstable nuclei have been measured using exotic 
isotope beams produced by projectile fragmentation 
in high-energy heavy-ion collisions.  
An important motivation for these studies is to determine 
the size of unstable nuclei \cite{Tani85,Mitt87,Ozawa00}.  
These measurements have in fact discovered the largely extended 
structure of unstable nuclei such as $^{11}$Li  \cite{Tani85}, 
$^{11}$Be \cite{Fuk91}, and $^{17,19}$C~\cite{Ozawa00}.
Nuclei 
with an extended density distribution are referred to as ``halo'' nuclei 
and their structure has been studied extensively as one of the characteristic 
features of weakly bound nuclei. 
A proton halo has also been claimed 
for $^{8}$B 
due to 
a large enhancement of the quadrupole moment~\cite{Minami93}.  
The root-mean-square radius diverges for $s$ and $p$ waves as
 the single-particle energy approaches zero ~\cite{Riisager} 
and the halo structure
can be ascribed to an occupation of an $l=0$ or $l=1$ orbit by the valence 
nucleons near the threshold~\cite{Sagawa93}.  

Recently, Takechi {\it et al.}
measured the 
interaction cross sections of neutron-rich Ne isotopes $^{28-32}$Ne
~\cite{Takechi10} 
and reported a halo structure of the $^{31}$Ne nucleus. 
The halo structure of  $^{31}$Ne was 
also reported in the one-neutron removal reaction 
by Nakamura {\it et al.}~\cite{Naka09}.
The interaction cross section $\sigma_I$ 
is almost the same as 
the reaction cross section 
$\sigma_R$ for unstable nuclei since the 
cross sections for inelastic scattering 
are in general small \cite{OYS92,KIO08}. 
In addition to 
large reaction cross sections for neutron-rich Ne isotopes, 
the experimental data also show that 
the reaction cross sections of $^{30}$Ne and  $^{32}$Ne are quenched 
as compared to the neighboring odd Ne isotopes, that is, 
a clear odd-even staggering has been found in the reaction cross 
sections~\cite{Takechi10}. 

In this paper, we discuss the relation between the odd-even staggering 
in the reaction cross sections and the so called pairing anti-halo 
effect \cite{Benn00}.  
The pairing correlation 
has been known important 
in nuclei throughout the periodic table, 
giving an extra binding for paired nucleons~\cite{BM75,BB05}.  
It plays an essential role especially in 
loosely bound nuclei such as $^{11}$Li and $^{6}$He (so called the 
Borromean nuclei), because the nuclei will be unbound 
without the pairing correlations. 
It is known that many-body correlations strongly modify 
the pure mean-field picture of loosely bound nuclei. It may happen that   
the strong pairing correlations do not allow 
the decoupling of weakly bound nucleons from the core nucleus and prevent
the growth of halo structure.  
This pairing ``anti-halo'' effect has been suggested 
with the Hartree-Fock Bogoliubov (HFB) method ~\cite{Benn00}, 
but a clear experimental signature of the anti-halo effect has not yet 
been obtained so far. 
We will examine in this paper whether
the odd-even staggering observed in the recent experimental data for the 
reaction cross sections of neutron-rich Ne isotopes 
is an evidence for the pairing anti-halo effect. 

In order to address this question, 
we first 
study neutron-rich Ne isotopes using the HFB method. 
The asymptotic behavior of a Hartree-Fock wave function for $s-$wave 
reads 
\begin{equation}
 \psi(r) \sim \exp(-\alpha r),
\end{equation}
where $\alpha=\sqrt{2m|\varepsilon|/ \hbar ^2}$ with the HF energy
$\varepsilon$.
The mean square radius of this wave function is then determined as  
\begin{equation}
 \langle r^2\rangle_{\rm HF}=\frac{\int r^2 |\psi(r)|^2 d{\bf r}}
{\int  |\psi(r)|^2 d{\bf r}} \propto 
   \frac{1}{\alpha^2}= \frac{\hbar^2}{2m|\varepsilon|},
\end{equation}
which will diverge in the limit of vanishing separation energy $|\varepsilon|
 \rightarrow 0$.  
This divergence will occur not only for $s-$wave but also 
for $p-$wave, although the dependence on $|\varepsilon|$ is 
different\cite{Riisager}.  
In contrast, the upper component of a HFB wave function, which is 
relevant to the density distribution, behaves \cite{doba} 
\begin{equation}
  v(r) \propto \exp(-\beta r), 
\end{equation}
where $\beta$ is proportional to the square root of 
 the quasi-particle energy $E$,
\begin{equation}
\beta =\sqrt{\frac{2m}{\hbar ^2}(E-\lambda)},  
\end{equation}
$\lambda$ being the chemical potential. 
If we evaluate the quasi-particle energy 
in the BCS approximation 
or HFB with canonical basis, it is given as 
\begin{equation}
E=\sqrt{(\varepsilon -\lambda)^2 +\Delta^2}, 
\end{equation}
where 
$\Delta$ is the pairing gap.
For a weakly bound single-particle state with 
 $\varepsilon \sim 0$ and $\lambda  \sim 0$, the asymptotic 
behavior of the wave function 
$ v(r)$ is therefore determined by the gap parameter as, 
\begin{equation}
 v(r)\propto  \exp\left(-\sqrt{\frac{2m}{\hbar^2}\Delta}\cdot r\right).
\label{HFB}
\end{equation}
The radius of the HFB wave function will then 
be given in the limit of small separation energy $|\varepsilon|
 \rightarrow 0$ as 
\begin{equation}
 \langle r^2\rangle_{\rm HFB}
=\frac{\int r^2 |v(r)|^2 d{\bf r}}{\int  |v(r)|^2 d{\bf r}} \propto 
   \frac{1}{\beta^2} \rightarrow \frac{\hbar^2}{2m\Delta}.  
\end{equation}
As we show, the gap parameter  $\Delta$ stays finite even in the zero energy 
 limit of $\varepsilon$ with a density dependent pairing interaction.
Thus the extremely large extension of a halo wave function 
in the HF field will be reduced substantially by the pairing field 
and the root-mean-square (rms) 
radius of the HFB wave function will not diverge. 
This is called the 
anti-halo effect due to the pairing correlations \cite{Benn00}.

\begin{figure}
\includegraphics[scale=0.6,clip]{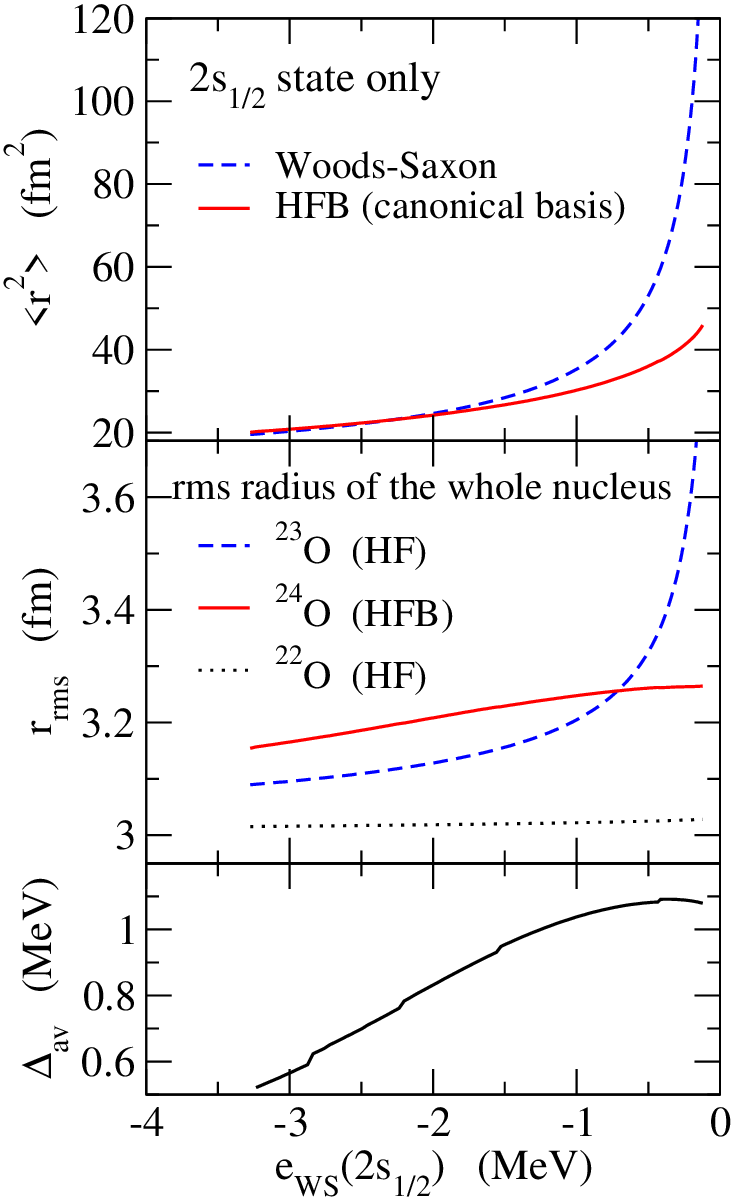}
\caption{(Color online) 
The mean square radii and the average paring gap as a function of 
the single particle energy $\varepsilon_{\rm WS}$ in a Woods-Saxon mean-field 
potential. 
The top panel shows the mean square radius of the 2s$_{1/2}$ wave function 
with and without the pairing correlation, denoted by 
HFB and Woods-Saxon, respectively.
The middle panel shows the rms radii 
for $^{22}$O (the dotted line), $^{23}$O (the dashed line), 
and $^{24}$O (the solid line), obtained with the Hartree-Fock ($^{22}$O and 
$^{23}$O) and the Hartree-Fock-Bogoliubov ($^{24}$O) calculations. 
The bottom panel shows the 
results of the HFB calculations for the 
average paring gap of $^{24}$O. 
}
\end{figure}

\begin{figure}
\includegraphics[scale=0.6,clip]{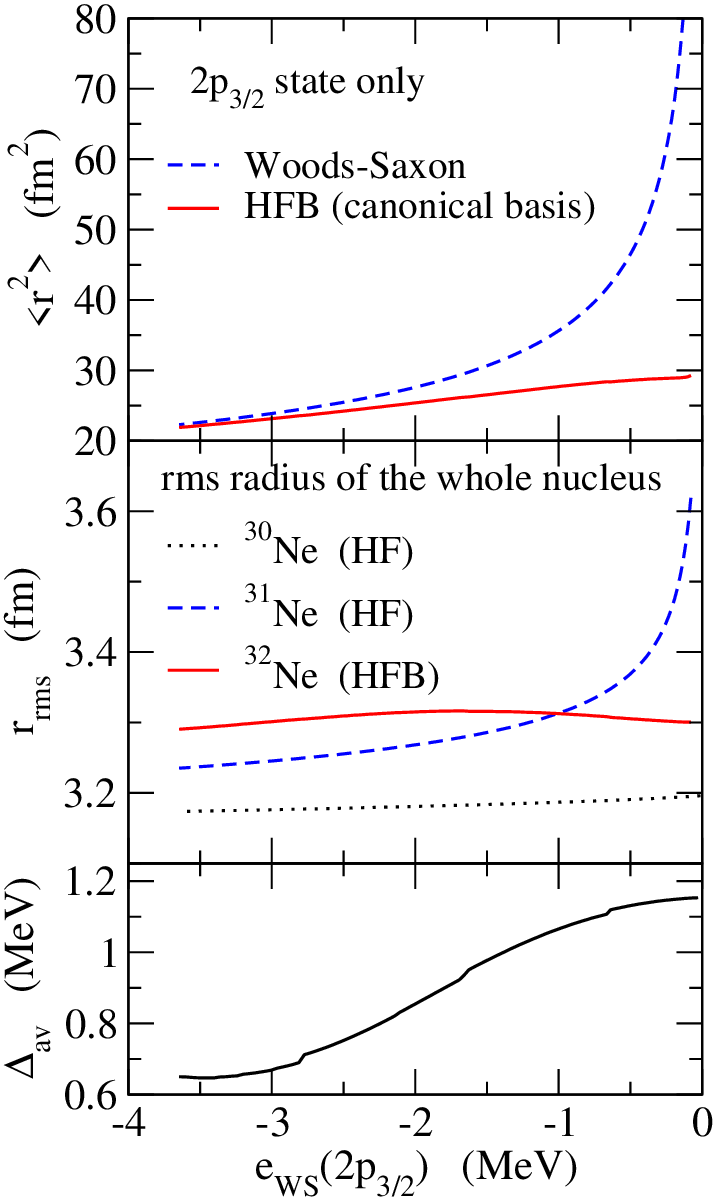}
\caption{ 
(Color online) Same as Fig.1, but for the 2$p_{3/2}$ state and for 
$^{30,31,32}$Ne isotopes. 
}
\end{figure}

We now numerically carry out 
mean-field calculations with a Woods-Saxon (WS) potential and also 
HFB calculations using the single-particle wave functions in the WS potential.
As examples of $s$-wave and $p$-wave states, we choose the 2$s_{1/2}$ state 
in $^{23}$O and 2$p_{3/2}$ state in $^{31}$Ne, respectively. 
Although $^{31}$Ne is most likely a deformed nucleus\cite{H10,UHS11}, for 
simplicity we assume a spherical Woods-Saxon mean-field potential. 
Notice that a Woods-Saxon potential with a 
large diffuseness parameter $a$ yields the 2$p_{3/2}$ state which is 
lower in energy than the 
1$f_{7/2}$ state, as was shown in Ref. \cite{HSCB10}. We use a similar 
potential with $a$=0.75 fm 
as in Ref. \cite{HSCB10} for $^{31}$Ne, while that in Ref. 
\cite{HS05} for $^{23}$O. 
For the HFB calculations, we use the density-dependent contact pairing 
interaction of surface type, in which the parameters are adjusted in 
order to reproduce the empirical neutron pairing gap for $^{30}$Ne \cite{YG04}. 
While we fix the Woods-Saxon potential for the mean-field part, the pairing 
potential is obtained self-consistently with the contact interaction. 

The top panel of Fig. 1 shows the mean square radius 
of the 2$s_{1/2}$ state for $^{23}$O, 
while that of Fig. 2 shows the mean square radius of the 
2$p_{3/2}$ state for $^{31}$Ne. 
In order to investigate the dependence on 
the single-particle energy, we vary the depth of the Woods-Saxon wells 
for the $s_{1/2}$ and $p_{3/2}$ states for $^{23}$O and $^{31}$Ne, respectively. 
The dashed lines are obtained with the single-particle wave functions, while 
the solid lines are obtained with the wave function for the canonical 
basis in the HFB calculations. 
One can see an extremely large increase of the radius 
of the WS wave function for both the $s$-wave and $p$-wave states 
in the limit of $\epsilon_{\rm WS}\rightarrow 0$. In contrast, 
the HFB wave functions show only a small increase of radius 
even in the limit of $\epsilon_{WS}\rightarrow 0$.  
This feature remains the same even when 
the contribution of the other orbits are taken into account, as shown 
in the middle panel of Figs. 1 and 2. 
Due to the pairing effect in the continuum,
the HFB calculations yield a larger radius than the HF calculations 
for the cases of $\epsilon_{\rm WS}\le -1$ MeV.  
On the other hand, in the case of
$-1$ MeV $<\epsilon_{\rm WS}<0$ MeV 
the HF wave function 
(equivalently one quasi-particle wave function in HFB) extends largely, 
while the HFB wave function does not get much extension 
due to the pairing anti-halo effect.  

In the bottom panel of Figs. 1 and 2, the average pairing gaps
are shown as a function of the single particle energy $\epsilon_{\rm WS}$.  
It is seen that
the average pairing gap increases 
as $\epsilon_{\rm WS}$ approaches zero. 
See also Fig. 2 in Ref. \cite{Benn00}. 
This is due to the 
fact that the paring field couples with the extended wave functions 
of weakly bound nucleons in the self-consistent calculations.  
That is, the pairing field is 
extended as the wave functions do and becomes larger for a loosely bound 
system.  
On the other hand, the paring gap behaves differently in the 
non-self-consistent model for the pairing field, 
in which the pairing field is fixed while solving the HFB equations
~\cite{Ham05}.  
In this model, 
the pairing field has a smaller overlap with the loosely bound nucleons 
because of the lack of the coupling, leading to 
a smaller paring gap in the limit $\epsilon\rightarrow 0$.  
Thus the anti-halo effect is somewhat underestimated in the 
non-self-consistent model as 
is expected from the asymptotic behaviour of the HFB 
wave function given by Eq. (\ref{HFB}). 

\begin{figure}
\includegraphics[scale=0.45,clip]{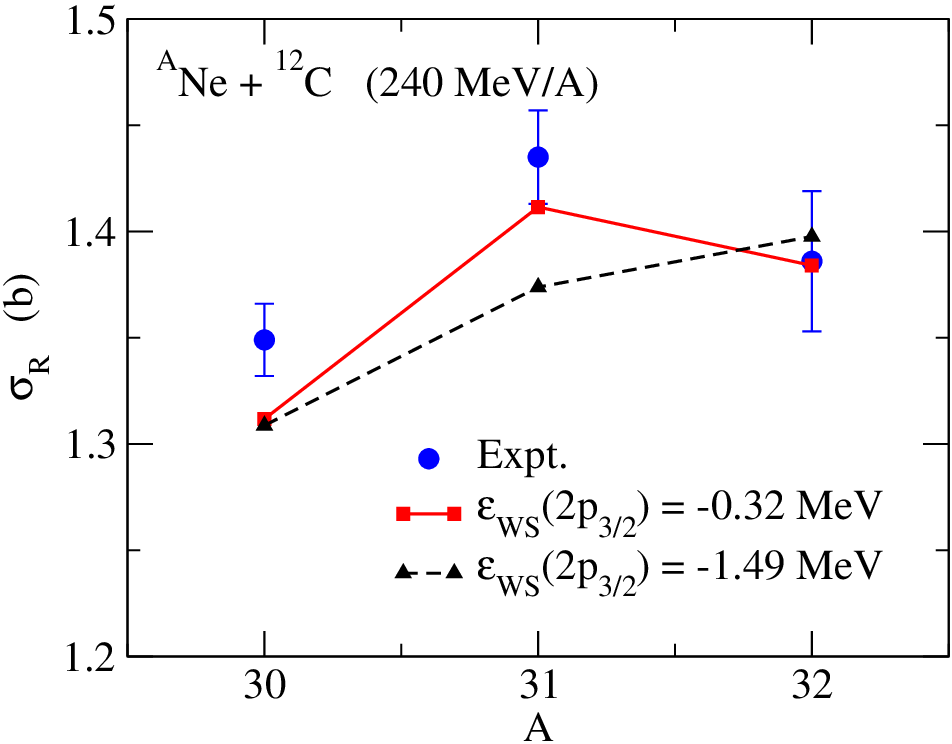}
\caption{(Color online) 
Reaction cross sections of Ne isotopes on a $^{12}$C target at 
$E_{\rm lab}$=240 MeV/A.  
The cross sections are calculated with the 
Glauber theory with HF and HFB densities. 
The solid line with the filled squares shows the results 
of $S_n$($^{31}$Ne) =$|\varepsilon (2p_{3/2})|$=0.32 MeV, 
while the dashed line with 
the open triangles is obtained 
for $S_n$($^{31}$Ne) =$|\varepsilon (2p_{3/2})|$=1.49 MeV. 
The experimental data are taken from Ref. \cite{Takechi10}.
 }
\end{figure}

Let us now calculate the reaction cross sections for the $^{30,31,32}$Ne 
isotopes and discuss the role of pairing anti-halo effect. 
To this end, we use the Glauber theory, in which we adopt the prescription 
in Refs. \cite{HSCB10,AIS00} in order to take into account 
the effect beyond the optical limit approximation. 
Fig. 3 shows the 
reaction cross sections of the $^{30,31,32}$Ne nuclei on a $^{12}$C target 
at 240 MeV/nucleon. 
We use the target density given in Ref. \cite{OYS92} and the profile function 
for the nucleon-nucleon scattering given in Ref. \cite{AIHKS08}. 
In order to evaluate the phase shift function, we use the two-dimensional 
Fourier transform technique \cite{BS95}. 
The cross sections $\sigma_R$ shown in Fig. 3 are calculated 
by using projectile densities 
constrained to two different separation energies of 
the 2$p_{3/2}$ neutron state. 
The dashed line with triangles is 
obtained using the wave functions with the separation energy 
$S_n$ ($^{31}$Ne) =$|\epsilon_{\rm WS}|$=1.49 MeV, while the solid line with squares is calculated with the 
wave functions of $S_n$($^{31}$Ne)=$|\epsilon_{\rm WS}|$=0.32 MeV. The empirical separation energy of 
$^{31}$Ne has a large ambiguity with $S_n$=0.29$\pm1.64$ MeV\cite{J07}.  
The cross section $\sigma_R$ of $^{30}$Ne is already much larger than 
the systematic values of Ne isotopes with A$<$30.  On top of that, 
we can see a clear odd-even staggering in the results with the 
smaller separation energy, $S_n$=0.32 MeV, as much as in the experimental data,  
while almost no staggering is seen in the case of the larger 
separation energy, $S_n$=1.49 MeV.  
This difference is easily understood 
by looking at the anti-halo effect for $|\epsilon_{\rm WS}|\le1$MeV shown in the 
middle panel of Fig. 2.  
Recently, the effect of deformation of neutron-rich Ne
isotopes on reaction cross sections was evaluated using a deformed Woods-Saxon model \cite{Minomo11}.  
It was shown that the deformation is large as much as $\beta_2\sim 0.42$ 
in $^{31}$Ne 
and enhances the reaction cross section by about 5\%.  However, 
the calculated results 
did not show any significant odd-even staggering in $\sigma_R$ between $^{28}$Ne and $^{32}$Ne \cite{Minomo11}.  

We mention that 
the pairing anti-halo effect may appear also in lighter halo nuclei. 
In Borromean systems, only the three-body system as a whole is bound and 
none of the two-body subsystems is unbound.  In the case of $^{11}$Li nucleus, 
this idea is implemented in that $^{11}$Li is bound while 
$^{10}$Li and a di-neutron are unbound. 
That is, the pairing correlations
work in the continuum and gain the binding energy 
to make the three-body system bound \cite{BE91,Zhukov93}. 
We can make a simple estimate on how large the anti-halo effect is  
in the $^{11}$Li nucleus. 
The experimental matter radii were obtained for $^{9}$Li and $^{11}$Li as  
\begin{eqnarray}
\sqrt{\langle r^2\rangle_m}
=2.43\pm0.02 \mbox{fm}\,\,\,\,\,\,  ^{9}\mbox{Li},  \nonumber \\
\sqrt{\langle r^2\rangle_m}=3.27\pm0.24 
\mbox{fm}\,\,\,\,\,\,\,\, ^{11}\mbox{Li}, 
\end{eqnarray}
respectively\cite{Tani85}. 
The matter radius of the two neutrons outside the $^{9}$Li core can be 
estimated by using a formula (see {\it e.g.}, Ref. \cite{Esbensen01}), 
\begin{equation}
\langle r^2\rangle_{\rm 2n}=\frac{11}{2}
\left(\langle r^2\rangle_m(^{11}\mbox{Li})-\frac{9}{11}
\langle r^2\rangle_m( ^{9}\mbox{Li})\right)
=(5.68)^2 \mbox{fm}^2.
\end{equation}
On the other hand, the rms radii of single particle states 
1$p_{1/2}$ and 2$s_{1/2}$ 
could be calculated with a Woods-Saxon potential 
adjusting the separation energy as $S_n=|\varepsilon_{\rm WS}|$=0.15 MeV;
 \begin{eqnarray}
\sqrt{\langle r^2\rangle }=6.48 \mbox{~fm} \,\,\,\,\,\, \mbox{for 1p}_{1/2} \nonumber \\
\sqrt{\langle r^2\rangle}=10.89 \mbox{~fm} \,\,\,\,\,\, \mbox{for 2s}_{1/2}.
\end{eqnarray}
These values lead to the rms radius of $^{11}$Li to be 3.53 fm and 5.14 fm 
assuming pure $(1p_{1/2})^2$ and $(2s_{1/2})^2$ two-neutron configurations  
outside the $^{9}$Li core, respectively. 
These values are apparently larger than the empirical value, 3.27 fm, 
and thus one sees the pairing 
anti-halo effect in the matter radius of $^{11}$Li.

\begin{figure}
\includegraphics[scale=0.45,clip]{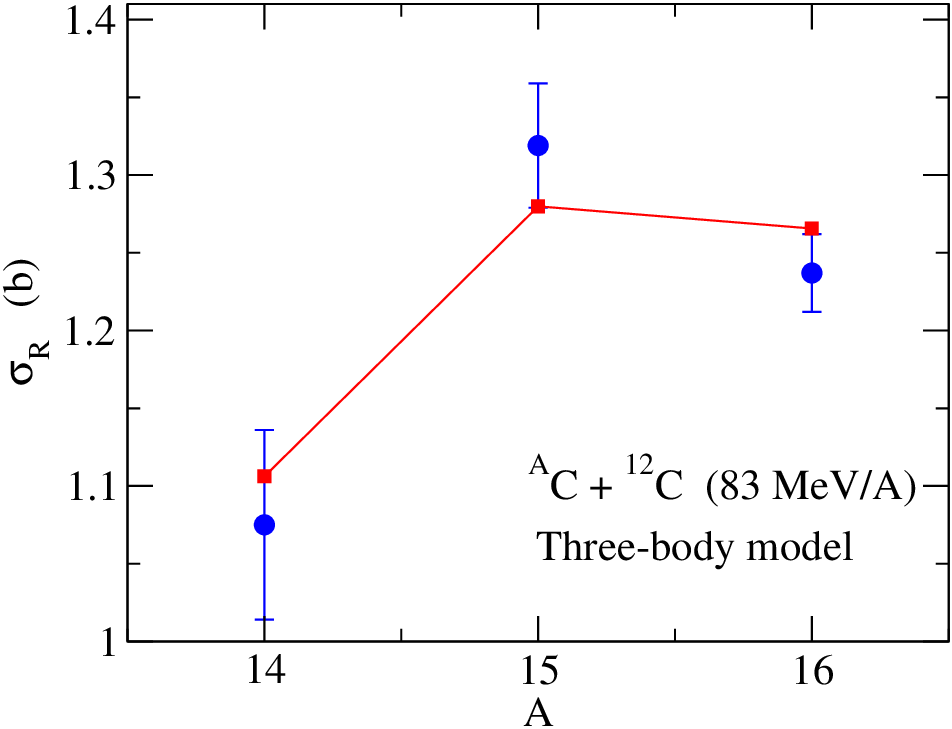}
\caption{(Color online) 
Reaction cross sections of C isotopes on a $^{12}$C 
target at $E_{\rm lab}$=83 MeV/A.  
The cross sections are calculated with the 
Glauber theory with three-body model densities. 
The experimental data are taken from Ref. \cite{FYZ04}.
 }
\end{figure}

The manifestation of the pairing anti-halo effect may be trivial 
for $^{11}$Li as $^{10}$Li is unbound. We therefore investigate next 
neutron-rich C isotopes. We particularly study the $^{16}$C nucleus 
using a three-body model given in Ref. \cite{HS07}. 
In this case, the valence neutron in $^{15}$C occupies the 2$s_{1/2}$ 
level at $\epsilon_{\rm WS}=-1.21$ MeV, while 
$^{16}$C is an admixture of mainly the $(2s_{1/2})^2$ and 
$(1d_{5/2})^2$ configurations. 
Assuming the set D given in Ref. \cite{HS07} 
for the parameters of the Woods-Saxon 
and the density distribution for $^{14}$C given in 
Ref. \cite{PZV00}, the rms radii are estimated to be 
2.53, 2.90, and 2.81 fm for $^{14}$C, $^{15}$C, and $^{16}$C, 
respectively. 
The corresponding reaction cross sections $\sigma_R$ calculated with 
the Glauber theory are shown in Fig. 4. 
The calculation well reproduces the experimental odd-even staggering 
of the reaction cross sections, that is a clear manifestation of 
the pairing anti-halo effect. 

In summary, 
we have studied the mass radii of Ne isotopes with the Hartree-Fock (HF) and 
Hartree-Fock-Bogoliubov (HFB) methods with a Woods-Saxon potential. 
The reaction cross sections 
$\sigma_R$ 
were calculated using the Glauber 
theory with these microscopic densities. 
We have shown that the empirical odd-even staggering in the 
reaction cross sections of neutron-rich 
Ne isotopes with the mass A=$30\sim32$ is 
well described by the HFB density and can be 
considered as a clear manifestation of the pairing 
anti-halo effect associated with 
a loosely-bound 2$p_{3/2}$ 
wave function. 
The  anti-halo effect was examined by the three-body 
model calculations also for $^{11}$Li and $^{16}$C.  
We argued that the observed mass radius of $^{11}$Li can be considered 
as an evidence for the 
anti-halo effect. 
We have also shown that the experimental reaction cross sections 
of $^{14-16}$C are well reproduced, confirming  
the ani-halo effect in these nuclei.  

We point out 
that the odd-even staggering in reaction cross sections has been widely 
observed in a series of nuclei with small separation energies, 
such as $^{14-16}$C \cite{FYZ04}, 
$^{18-20}$C\cite{Ozawa00}, 
$^{28-30}$Ne, $^{30-32}$Ne\cite{Takechi10} 
and $^{36-38}$Mg\cite{Takechi11}. 
Evidently, 
the pairing anti-halo effect is commonly seen 
in those weakly-bound nuclei near the drip line.
 
\medskip

We would like to thank M.  Takechi and H. Sakurai for fruitful discussions on 
experimental data, and C.A. Bertulani for useful discussions on the 
Glauber theory. 
This work was supported by the Japanese
Ministry of Education, Culture, Sports, Science and Technology
by Grant-in-Aid for Scientific Research under
the program numbers  (C) 22540262 and  20540277.

\end{document}